\documentstyle[12pt,epsf]{article}

\newcommand{\be}{\begin{equation}}
\newcommand{\ee}{\end{equation}}
\newcommand{\bea}{\begin{eqnarray}}
\newcommand{\eea}{\end{eqnarray}}
\newcommand{\no}{\noindent}
\newcommand{\sign}{{\rm sign}}

\begin{document}

\centerline{\bf STATISTICAL FEATURES IN LEARNING}\par\bigskip

\centerline{Ion-Olimpiu Stamatescu}\par\medskip

\centerline{\it Inst. Theor. Phys. der Universit\"at, 
Philosophenweg 16, D-69120 Heidelberg,}\par
\centerline{\it and}\par
\centerline{\it FEST, Schmeilweg 5, D-69118 Heidelberg, 
Germany}\par\bigskip

\noindent ABSTRACT: We study some features of learning models based
on ``delayed" and undifferentiated reinforcement. We concentrate on models  based on primitive 
algorithms without sophisticated structure and which might be considered as very elementary.

\section{Introduction}

 Introducing biologically motivated features in
models for learning has usually a double role: testing hypotheses  
for natural learning and finding hints for artificial learning. Here we 
do not take the more ambitious point of view of finding optimal algorithms for the latter.
On the contrary, our motivation is to investigate which are the capabilities of
very  elementary 
mechanisms. 
We shall consider a number of 
learning aspects which may appear biologically motivated (by which we mean
 primarily the functional aspects). Specifically we shall consider learning
models called here, for short, ``stochastic learning", as
involving the following features:
\begin{itemize}
\item [1.] implementation achieved by acting stochastically in a structured environment;
\item [2.]  control 
by non-specific reinforcement, that is, the reinforcement is given
according to the success of series of actions;
\item [3.]  reinforcement acting via the normal, internal activity
of the system (requires no ``external computation").
\end{itemize}

The first point is essential for the perspective taken here which sees 
learning as result of the interaction between an  outward action,
spontaneous and random in its basis, 
and an inward feedback reflecting the environmental conditions.
The second point represents the normal situation for this problem setting, since the agent usually will only find out at the end of the day whether he is 
successful or not (e.g., survives or not) -- but will not be told how good 
or bad was each of its steps. The third point is somewhat subtle,
essentially it suggests that no algorithm should be used which is of a higher
level than that defining the agent.

From the point of view of reinforcement learning our problem may be seen under 
``class III" problems in the classification of Hertz et al. (1991).   
However, we stress that our attitude is not that of finding good algorithms for tackling 
special problems, like movement, control or games -- see, e.g., 
Kaelbling (1996). Instead we want to test whether under these quite restrictive conditions the internal structure of the agent will become differentiated 
enough to meet the conditions of the environment. For this reason we 
do not consider evolved algorithms from the class of
 Q-learning (Watkins 1989), of TD learning (Sutton 1988), agent and 
critic (Barto, Sutton and Anderson 1983), etc but restrict to most primitive
 algorithms which we may think of having a chance to have developed 
under natural conditions. On the other hand, if such algorithms will
prove capable of tackling the problem they may well give further insights.

An illustration of the problem was provided in an earlier paper 
(Mlodinow and Stamatescu 1985) dealing with 
these questions in the simulation of a device
moving on a board.  The device realizes
 a biased random walk, the bias coming from trying to
recognize situations and considering the ``goodness"   of the previously
taken, corresponding moves. There is absolutely no structure presupposed
in the behavior of the device, beyond the sheer urge to move
(completely at random to start with), the structure is fully
hidden in the environment -- according to point 1. above. The
reinforcement (positive or negative) itself is global, it is associated 
to the results of  long
chains of moves. It is assigned undiscriminately to all  
moves in the chain to modify their ``degree of goodness" (see point 2.). Under these conditions the device shows a 
number of interesting, quantifiable features: ``flexible stability"
(its behavior fluctuates around a solution -- path to the goal -- without loosing it, unless as a result of the
fluctuations
a better solution is found);  ``development"
(in the course of the
training on harder and harder problems, solutions to the simpler problems are
taken over and applied to subproblems 
of the complex case); ``alternatives handling" 
(in a continuously changing environment the device
develops alternative solutions, which it distinguishes by simple cues
found
in the environment); ``learning from success and failure" (all
experiences contribute); etc. 

This simulation produced therefore arguments for the realizability of this
model of ``stochastic learning". However the last point  (3.) appeared less
clear, since the reinforcement, although simple, did not proceed
via the activity of the system itself: it had to {\it fit} the situations seen
to what it has saved,  it had to choose actions, it was given the 
possibility of recognizing impossible actions, etc. 

A more natural frame for our problem is indeed provided by neural networks.
In section 3 we shall present as illustration 
an analog simulation to the one mentioned above, realized
by a neural network, and where we conform in a clear way also to point 3. above.
However, in the context of neural networks we can  analyze such problems 
 in a more systematic way trying to achieve results of statistical significance, and not only illustratively. In the  section 2 we shall 
deal with a  realization of our problem on perceptrons.

\section{Learning rule for perceptrons under unspecific reinforcement}

We consider perceptrons with Ising units $s,s_i$ = 1,-1 
and real weights (synapses) $C_i$:
\begin{equation}
     s = \sign \left( \sum_{i=1}^N C_i s_i\right) = 
         \sign \left( {\bf C}.{\bf {s}}\right)
\label{e.1}
\end{equation}
\no $N$ is the number of neurons and we put no explicit thresholds. 
The network (pupil) is presented with a series of patterns 
$s_i^{(q,l)}$,  $q=1, ... ,Q$, $l=1, ... ,L$
to which it answers with $s^{(q,l)}$. A training period consists of
the successive presentation of $L$ patterns.
The answers are compared with the corresponding answers  $t^{(q,l)}$
of a teacher
with pre-given weights $T_i$ and
the average error made by the pupil over one
training period is calculated:
\be
    e_q = \frac{1}{2L} \sum_{l=1}^L |t^{(q,l)} -s^{(q,l)}| .
\label{e.2}
\ee
\no The training algorithm
consists of two parts:\par
\begin{itemize}
\item [I.] - a ``blind" Hebb-type  {\it association} at each presentation of
a pattern:
\be
   {\bf C}^{(q,l+1)} ={\bf C}^{(q,l)} + a_1 s^{(q,l)}{\bf s}^{(q,l)};
\label{e.3}
\ee
\item [II.] - an ``unspecific" {\it reinforcement}, also Hebbian, at the end
of each training period
\be
   {\bf C}^{(q+1,1)} = {\bf C}^{(q,L+1)} - a_2 e_q
                \sum_{l=1}^L s^{(q,l)}{\bf s}^{(q,l)}.
\label{e.4}
\ee
\end{itemize}
\no We shall call this algorithm ``2-Hebb rule" and we shall be interested in the behavior of the error $e(Q)$ for large $Q$ -- or, alternatively, the approach of the pupil weights toward those of the teacher $P(Q)$:
\be
  e(Q) = {1 \over {2M}} \sum_{m=1}^M |t^{(m)} - \sign \left( {\bf C}^{(Q,1)}
.{\bf s}^{(m)}\right)|\ , \ \ 
   P(Q) = {{{\bf C}.{\bf T}} \over {|{\bf C}|~|{\bf T}|}}
\label{e.5}
\ee
\no Both the training patterns $s^{(q,l)}$ and the test patterns $s^{(m)}$
are generated randomly. We shall 
test whether the behavior of $e(Q)$ can be reproduce by a power law at large $Q$:
\be
      e(Q) \simeq {\rm const} Q^{-p}
\label{e.6}
\ee

 \no Notice the following features:
\begin{itemize}
\item [a)] In the training the pupil only uses its own associations {\bf s}$^{(q,l)}
, s^{(q,l)}$ and the average error marge $e_q$ which does not refer specifically to the particular steps $l$.
\item [b)] Since the answers $s^{(q,l)}$ are made on the basis of the instantaneous
weight values {\bf C}$^{(q,l)}$ which change at each step according to eq. (3), the series of 
answers form a correlated sequence with each step depending on the previous one.
Therefore $e_q$ measures in fact the performance of a ``path", an interdependent set of decisions.
\item [c)] For $L=1$ the algorithm reduces of course to the usual 
``perceptron rule" (for 
$a_1=0$) or to the usual ``unsupervised Hebb rule" (for $a_2 = 2 a_1$). We
have on general arguments $p=1$ for the first, $p=0.5$ for the second case.
\end{itemize}

\no In the following we shall present preliminary results from an on going project
(K\"uhn and Stamatescu 1998). Here $a_1, a_2$ are either fixed or  tuned 
according to the following rule:
\be
    a_1(q) = {a \over 2}e_q (e_q + {a \over 2}), \  \ 
    a_2(q) = a e_q .
\label{e.7}
\ee

\begin{figure}[tb]
\vspace{7cm}
\includegraphics{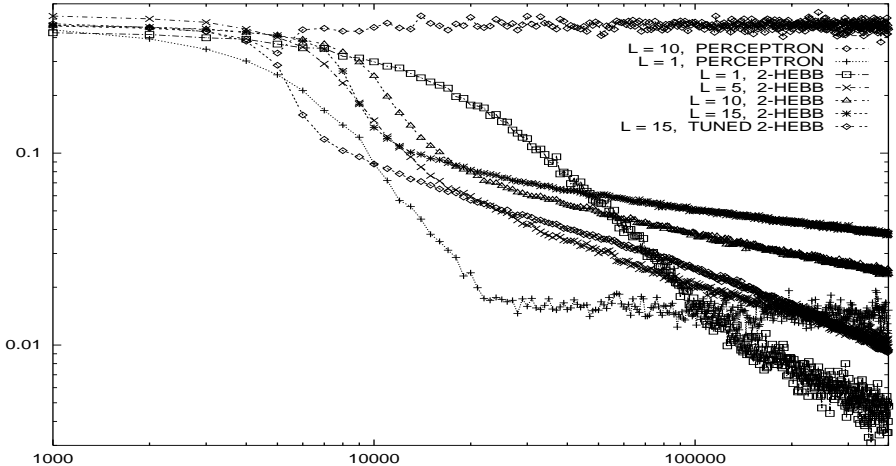}
\caption{{\it Error vs number of training periods for N = 200, various L. For 
the ``perceptron" rule $a_1 = 0.$, for the ``2-Hebb" rule 
$a_1 / a_2 : 0.2 - 0.25$. The ``tuning" is given by eq. (7) with $a=0.1/N$}.
}
\label{f.p200}
\end{figure}

\begin{figure}[tb]
\vspace{5.5cm}
\includegraphics{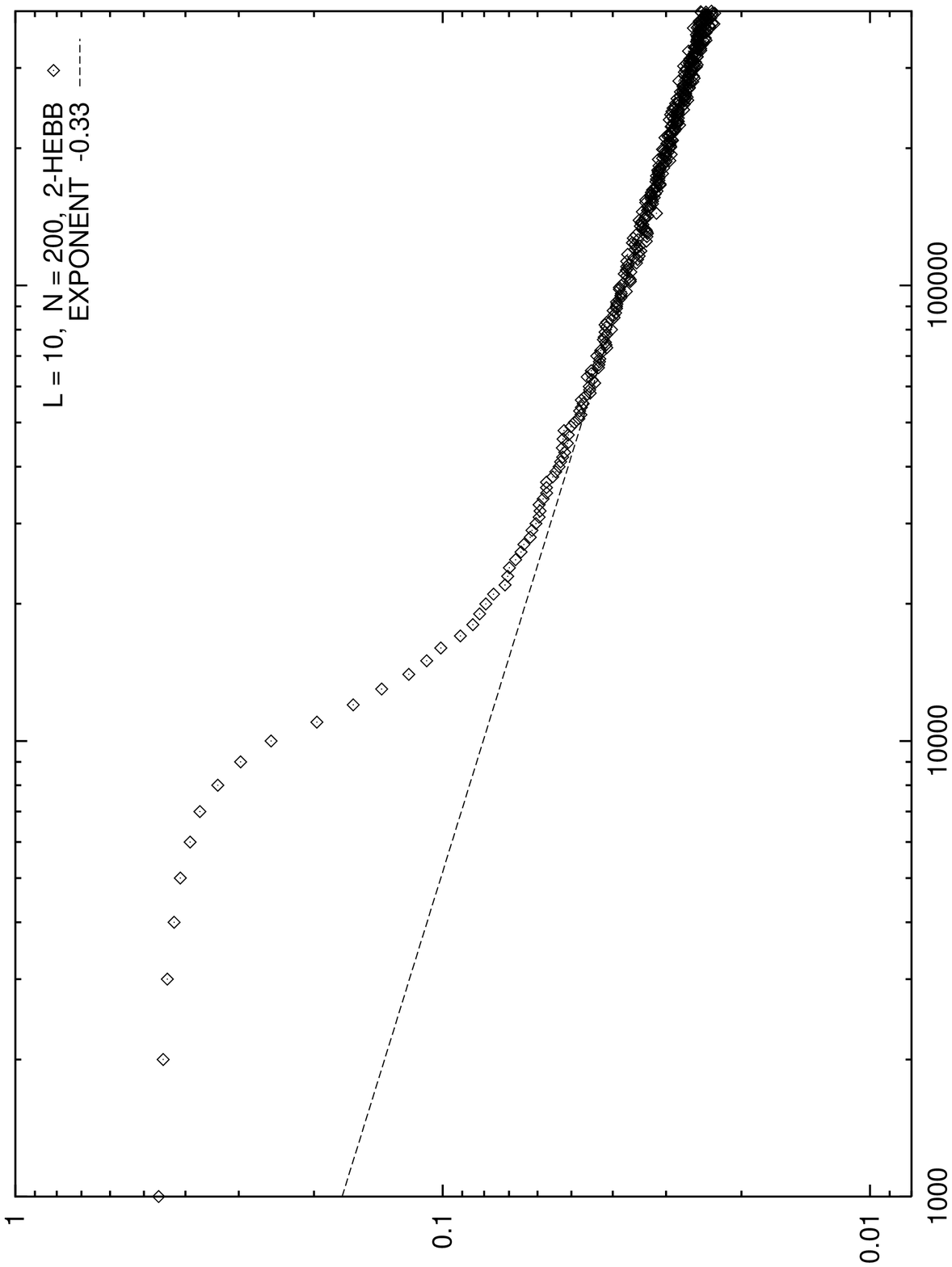}
\includegraphics{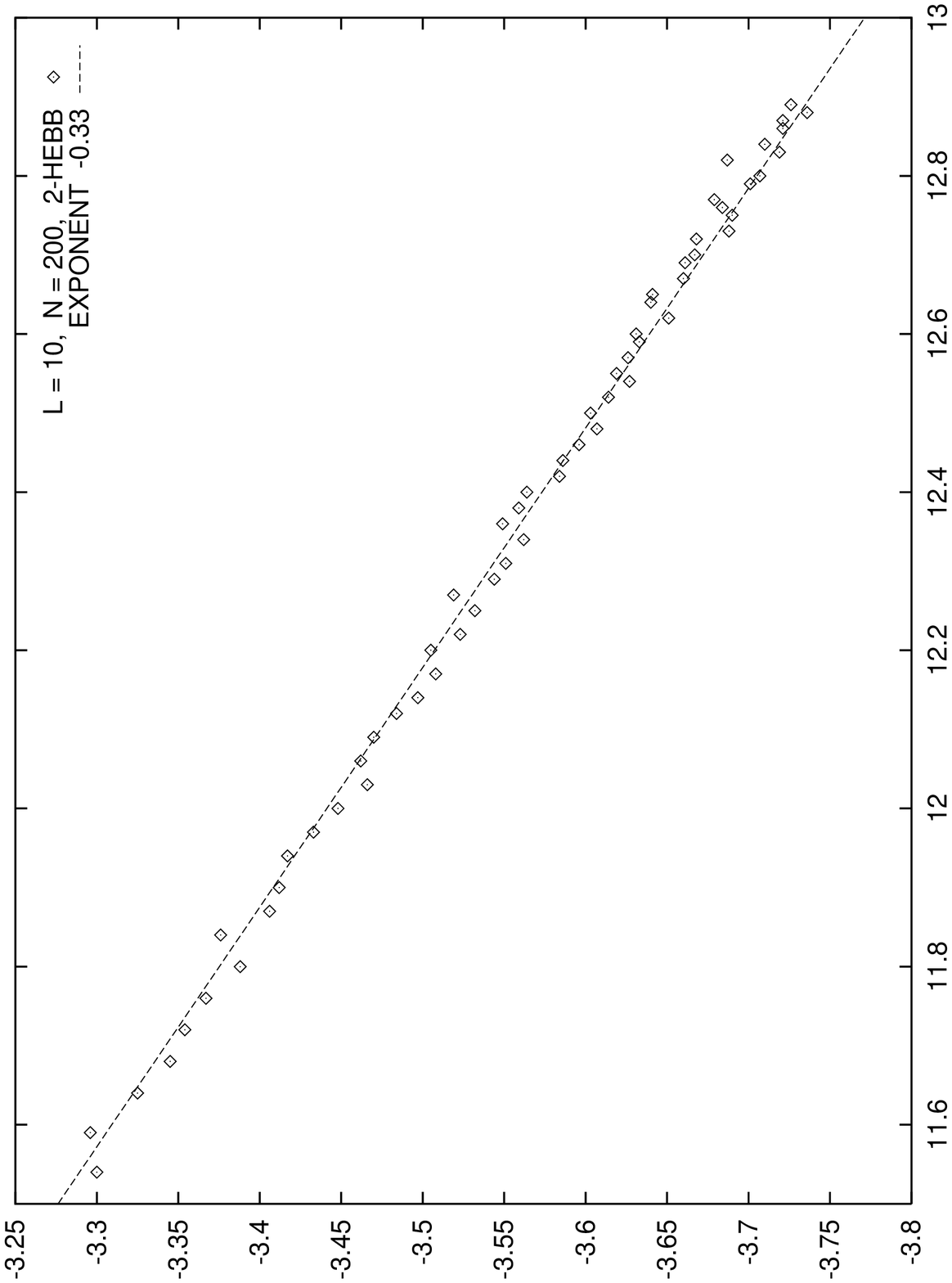}
\caption{{\it Error vs number of training periods for N = 200, L = 10, 
``2-Hebb" rule with $a_1 = 0.01/N$, $a_2 = 0.05/N$; $p=0.33$. The right hand
plot gives directly the logarithms in the ``asymptotic" regime}.
}
\label{f.p10}
\end{figure}

\begin{figure}[tb]
\vspace{5.5cm}
\includegraphics{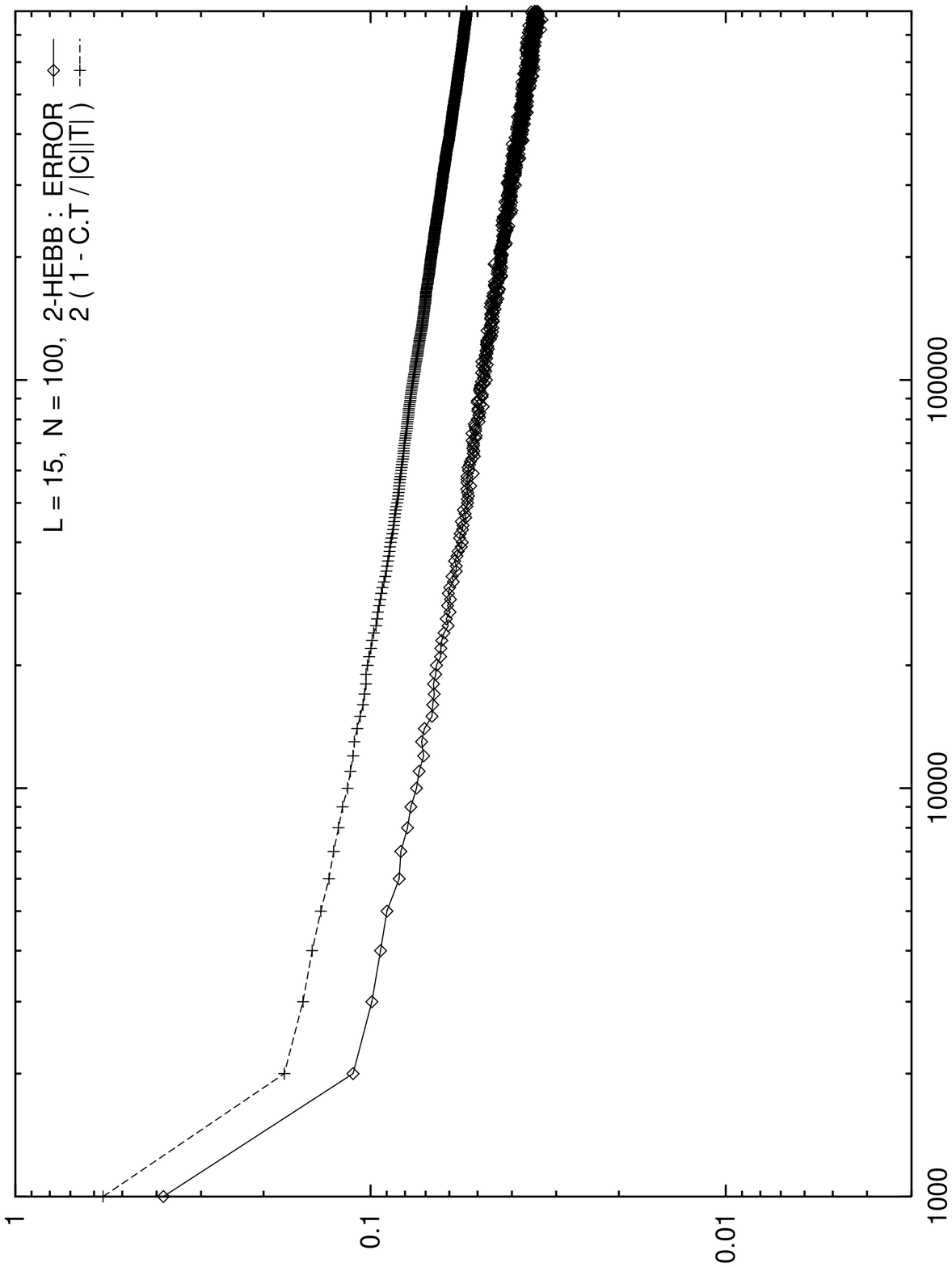}
\includegraphics{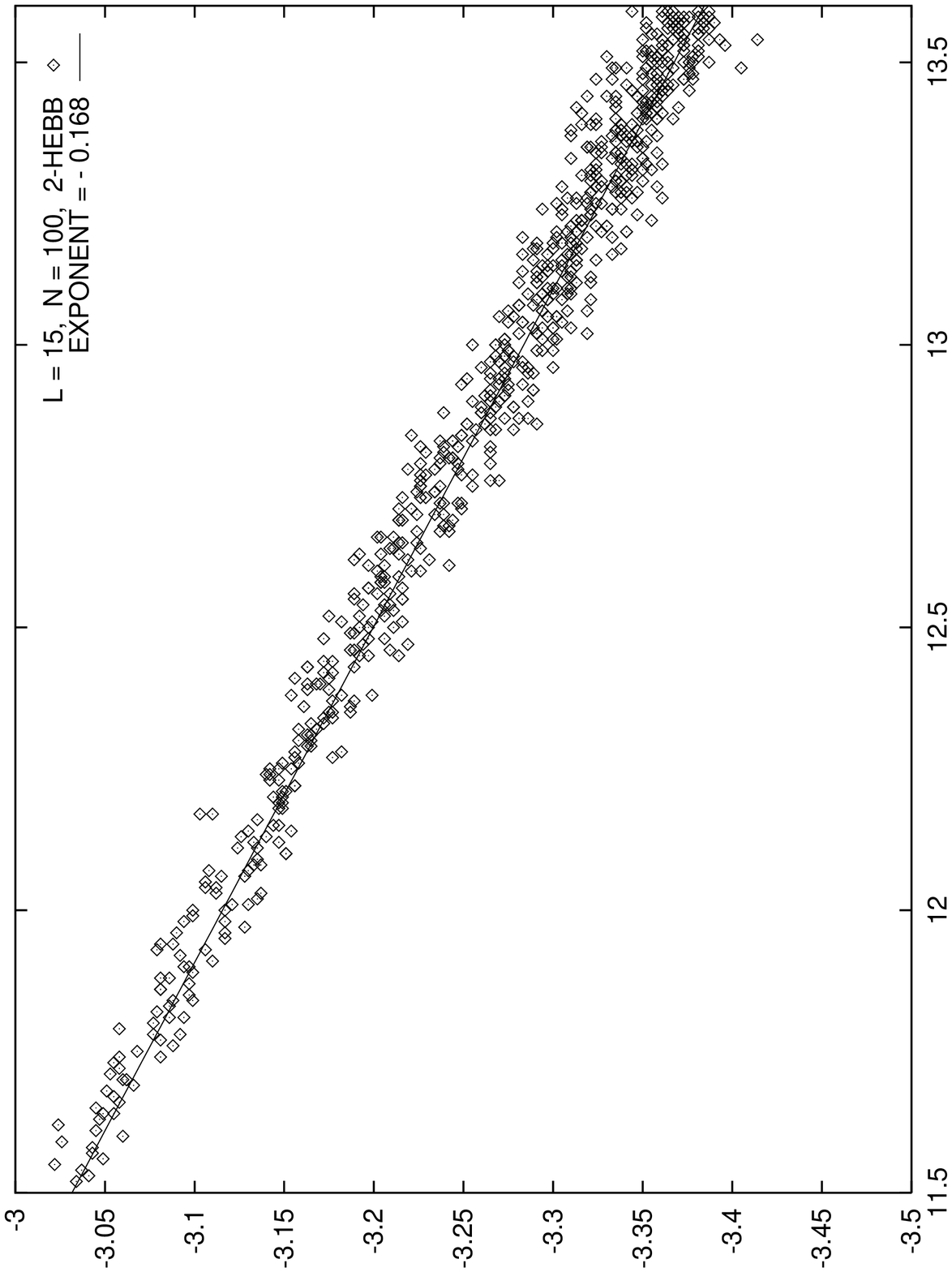}
\caption{{\it Error vs number of training steps for N = 100, L = 15, 
``2-Hebb" rule with $a_1 = 0.01/N$, $a_2 = 0.05/N$; $p=0.168$. The left hand
plot presents also the convergence of the weight ray -- see eq. (5). The right hand
plot gives directly the logarithms in the ``asymptotic" regime}.
}
\label{f.p15}
\end{figure}

\begin{figure}[tb]
\vspace{5.5cm}
\includegraphics{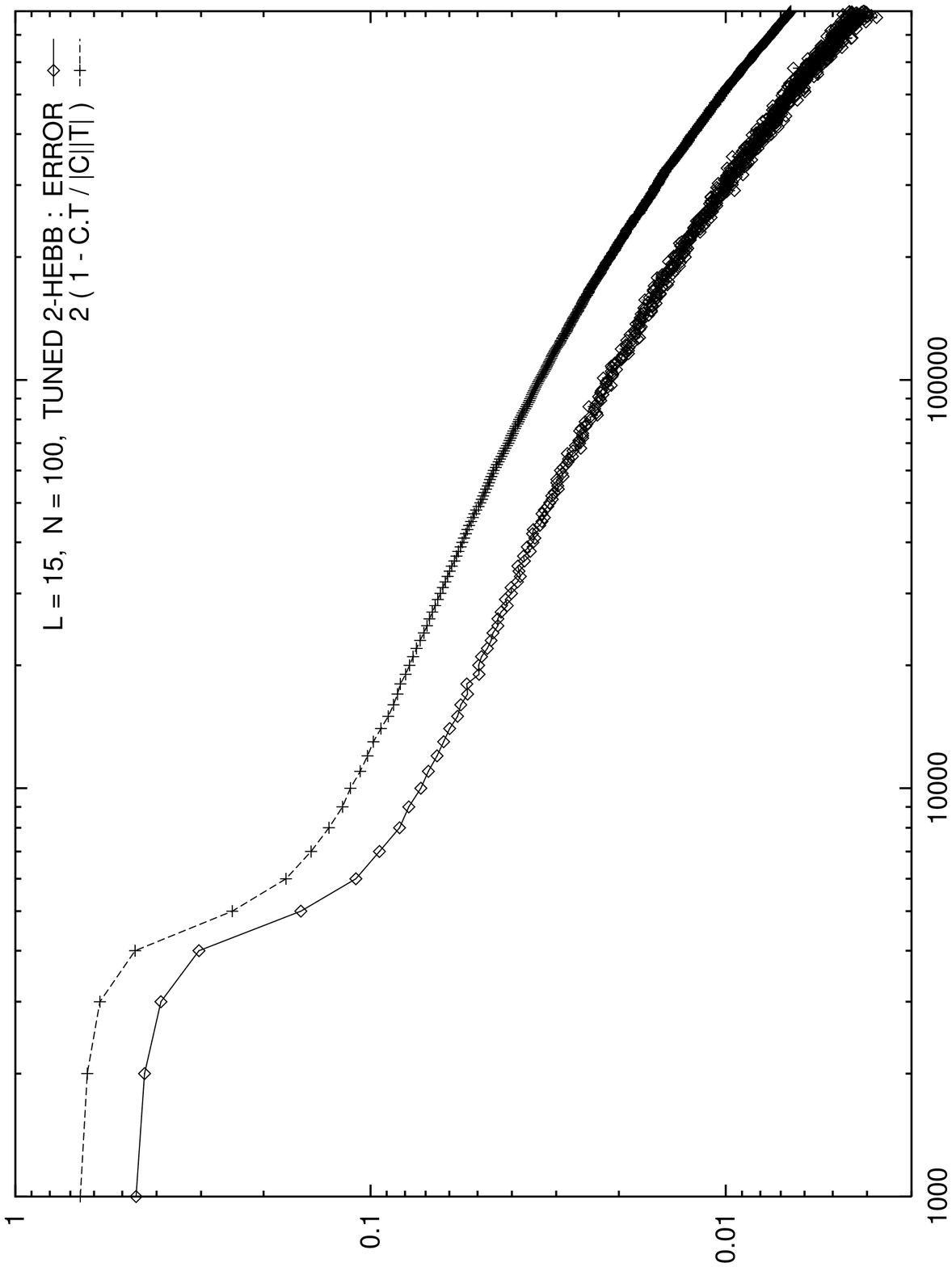}
\includegraphics{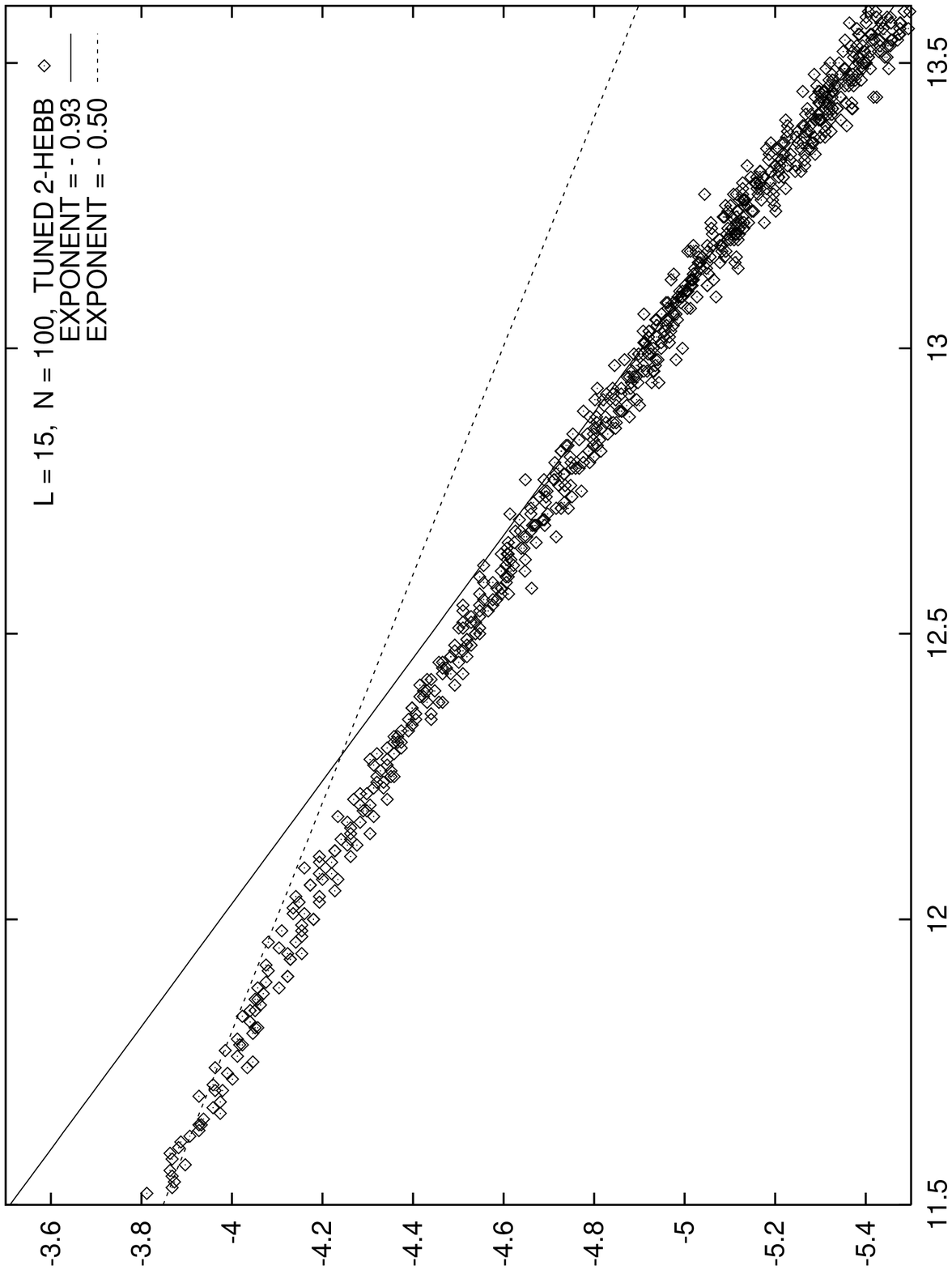}
\caption{{\it Error vs number of training periods for N = 100, L = 15, 
``tuned 2-Hebb" rule with $a = 0.1/N$, see eq. (7). 
The ``asymptotic" exponent is $p=0.93$. The left hand
plot presents also the convergence of the weight ray -- see eq. (5). The right hand
plot gives directly the logarithms in the ``asymptotic" regime}.
}
\label{f.m15}
\end{figure}

\no  We use $M = 10000$ and $Q$ up to $8.10^5$. We tested  various combinations
of $L =1, 5,10,15$ and $N=50,100,200,300$. The general observations are:
\begin{itemize}
\item [-] For fixed $a_1, a_2$ and for $L$ of 5 and higher there is a rather narrow
region of ratios $r = a_1/a_2$, namely $r$: 0.2 - 0.25 for which we have convergence, in particular we find no convergence for $a_1=0$. See Fig.
\ref{f.p200}. For $L=1$ varying $a_1$ interpolates between perceptron and Hebbian learning, we did not performed a systematic analysis for $L=1$, however.
\item [-] The asymptotic behavior with $Q$ seems to be well reproduced by a power law. For fixed $a_1,a_2$ the exponent appears to depend on $L$, however we
may not have yet achieved convergence even for $N=300$. See Figs. \ref{f.p10} 
and \ref{f.p15}. For $a_1,a_2$ tuned according to eq. (7) the asymptotic 
behavior approaches the simple perceptron ($p = 1$). See Fig. \ref{f.m15}.
The exponent results are summarized in Table 1. 
\end{itemize}
\no $P(Q)$ behaves very similarly to $e(Q)$. Introducing thresholds or noise changes 
the particular results but does not seem to modify the general picture.
Using smooth response functions appears also not to modify this picture, at
least as long as the non-linear character is preserved. A more detailed analysis will be presented elsewhere.

\begin{table}[ht]
\begin{center}
\label{t.expn}
\begin{tabular}{|c|c|c|c|c|c|}
\hline
 & L = 5&L = 10& L = 15& L = 15 (tuned)& L = 15 (tuned) \\ \hline 
N & \multicolumn{4}{c|}{$Q_{max} = 400000$} & $Q_{max} = 800000$ \\ \hline
50& 0.42(4)&0.23(1)&0.14(1)& 0.84(2)& \\ \hline
100 &0.43(3) &0.27(1)&0.18(2)&0.80(3)& 0.93(4)\\ \hline
200&0.52(4) &0.30(2)&0.19(1) &0.65(5)&  \\ \hline
300&        &      &0.18(1)&0.60(5)&     \\ \hline
\end{tabular}
 
\caption[]{{\it Exponents of the power law ansatz for the error dependence 
on the number of training steps.}}
\end{center}
\end{table}

Attempting to understand the above results we use a {\it coarse grained} 
description of the training process by introducing a mean weight vector
{\bf U}(q) averaging over a learning period. For the coarse grained vector 
{\bf U}(q) the updating reads:
\be
{\bf U}(q+1) = {\bf U}(q) + (a_1 - a_2 e_q) \sum_{l=1}^L 
s^{(q,l)} {\bf s}^{(q,l)}
\label{e.8}
\ee
\no Taking the scalar product with {\bf s}$^{(q,l')}$ and using the randomness of the patterns we obtain after further approximations
\be
z(q+1) - z(q) = - c \left(a_1 - {{a_2} \over 2}  z(q) \right) (1 - z(q))
\label{e.9}
\ee
\no where $c$ is some positive constant and $z(q)$  
is the coarse grained quantity associated to 
\be
x^{(q,l)} = ( t^{(q,l)} - s^{(q,l)} ) t^{(q,l)} = \{0,2\}
\label{e.10}
\ee
\no From eq. (9) we see that  asymptotically convergence requires 
\be
a_1 > {{a_2} \over 2}  z(q)  > 0 \ \ {\rm for}\ \ z(q) < 1
\label{e.11}
\ee
\no while at the beginning we may need
\be
a_1 < {{a_2} \over 2} z(q) \ \ {\rm for}\ \ z(q) > 1
\label{e.12}
\ee
\no This very rough analysis (which can only serve as orientation)
seems to agree unexpectedly
 well with the numerical results. The tuning eq. (7) has been chosen in
agreement with eq. (11). 

The following simple observation may help to understand these findings: 
since for $L=1$ $e_q$ can only take the values 0 and 1  
$a_1=0$ means penalty for failure, no change for success, i.e. the
usual perceptron learning rule known to converge. However, for $L>1$
$e_q$ can take fractional values in the interval $[0,1]$. In this case
$a_1=0$ means penalty for all answers which are short of perfect. A special
case may be $L=2$: then there is only one
intermediate value, $e_q=0.5$, meaning ``undecided", and putting a penalty
on it does not destabilize the system.

\begin{figure}[tb]
\vspace{5cm}
\includegraphics{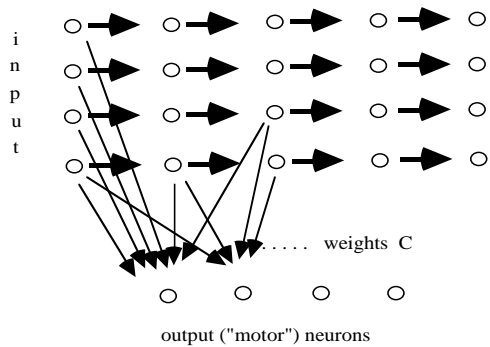}
\caption{{\it Architecture of the robot}.
}
\label{f.arti0}
\end{figure}

\section{A stochastic learning model in a (slightly) more complex situation}

In partial analogy with the model described in Mlodinow and Stamatescu (1985) we
consider a ``robot" moving on a board with obstacles under the following 
conditions:
\begin{itemize}
\item [1.] The board is realized as a grid and the robot can look and move
one step in each of the 4 directions (forth, right, back and left).
\item [2.] The robot's architecture consists of an input layer of $5 \times 4$
Ising neurons and an output layer of 4 (``motor") neurons, each responsible 
for moving in one direction. See Fig. \ref{f.arti0}. The modifiable synapses  are uni-directional
from all the input neurons to the output ones (120 weights). The output 
neurons are assumed to be interconnected such as to ensure
a ``winner takes all" reaction, with the winner being decided probabilistically
on the basis of the height of the presynaptic potentials.\footnote{The 
interconnection in the output layer has not been realized in the architecture and the
corresponding reaction has been put in  {\it by hand}, since this question
did not belong to the  ones we wanted to study -- it is
supposed the corresponding interconnection could be realized.}

\begin{figure}[tb]
\vspace{16cm}
\includegraphics{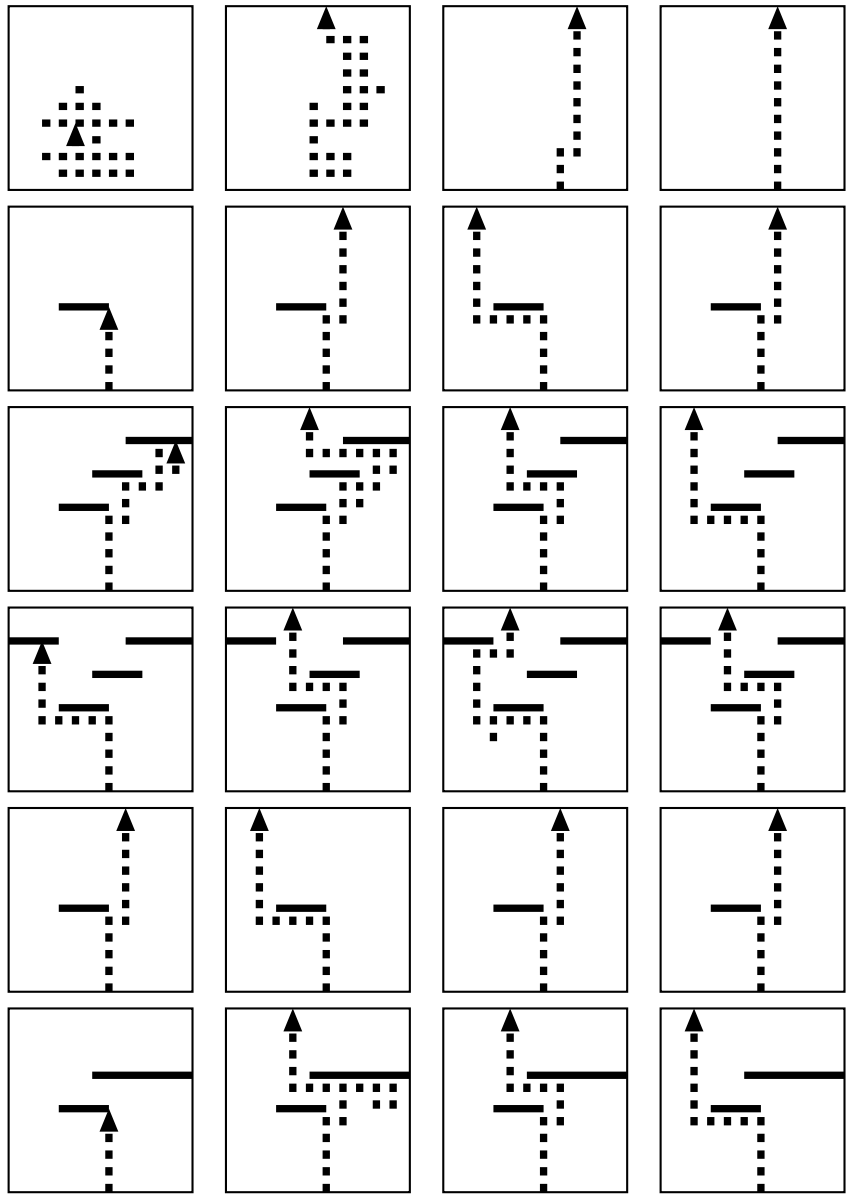}
\caption{{\it Typical results for 6 tests on which the robot is
trained consecutively (from up to down). From left to right: first run, early behavior, late behavior (including transients or alternatives)}.
}
\label{f.arti1}
\end{figure}

\item [3.] The immediate input is the state of the
4 neighboring cells (free-occupied); this information is always loaded in the 
the first 4 input neurons and transferred to the next group of 4 at the next step. Hence the robot has at each step as input the situations it has seen
at present and in the last 4 steps. For this input the presynaptic potentials of the motor neurons are calculated and a move is attempted.
\item [4.] The updating of the weights is achieved in 2 stages: \par
a. - At each step 
a ``blind" Hebb potentiation/inhibition is performed, considering only the 
actual states of the input and output neurons. If the prospected move 
runs against an obstacle the state of the output neuron responsible for 
this move is reversed before the Hebb updating is done.\par
b. - The robot starts from some given cell on the bottom of the board 
and moves. It stops if it arrives at the upper line of the board or if it had 
made some predefined, maximal number of steps.
The number of steps at stopping  
is compared with a predefined success marge and the difference 
is used to Hebb-inhibit/potentiate {\it equally} all synaptic updates it has
performed on the way (step a.).\par
\item [5.] The weights are not normalized; the various parameters (noise, 
thresholds, the amplitudes of the two updates) are given by best guess -- in a further development they may be left to the network itself to optimize (see 
Mlodinow and Stamatescu 1985).
\end{itemize}
\no The performance of the robot is illustrated in Figs. \ref{f.arti1} 
and \ref{f.arti2}.

\begin{figure}[tb]
\vspace{15cm}
\includegraphics{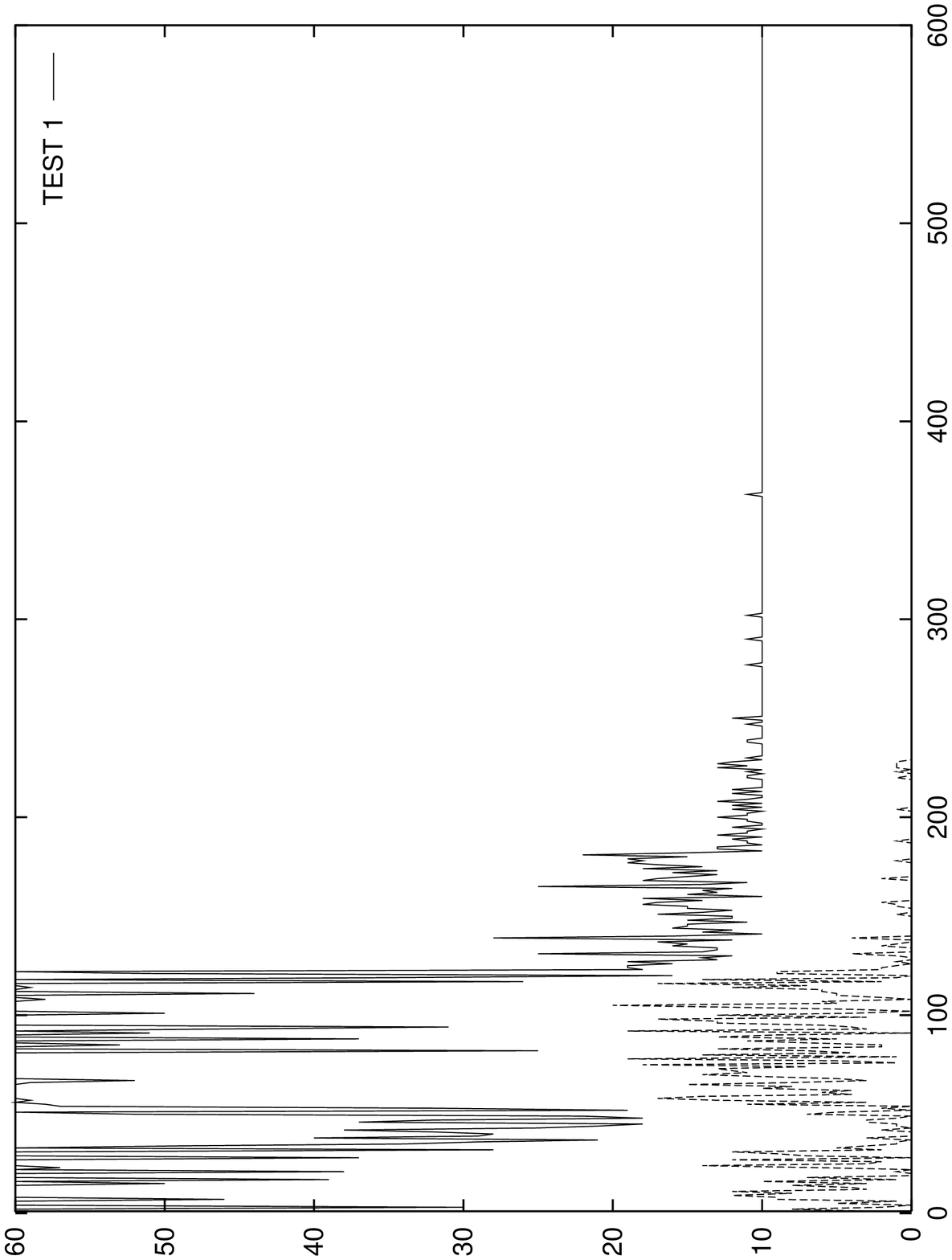}
\includegraphics{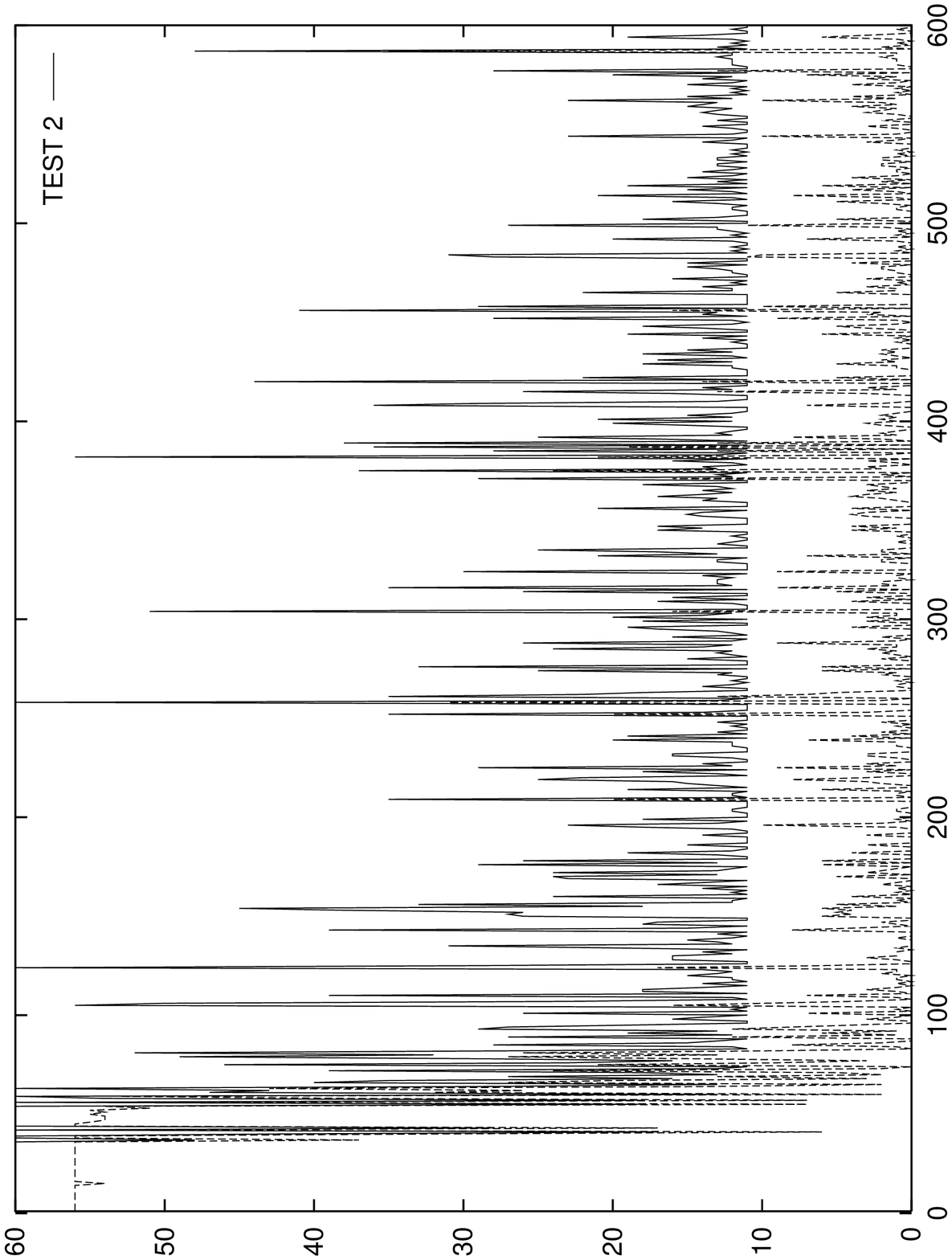}
\includegraphics{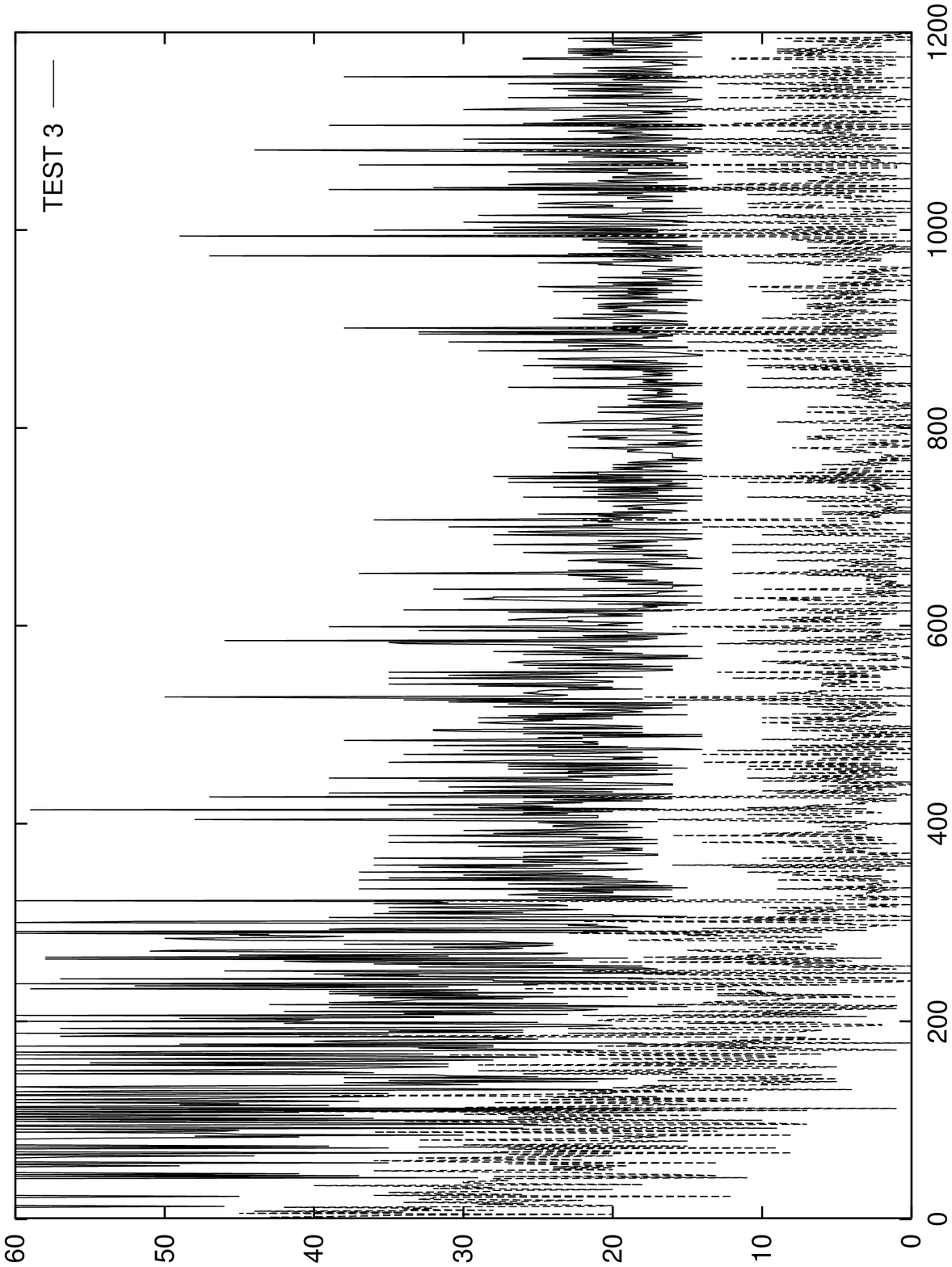}
\includegraphics{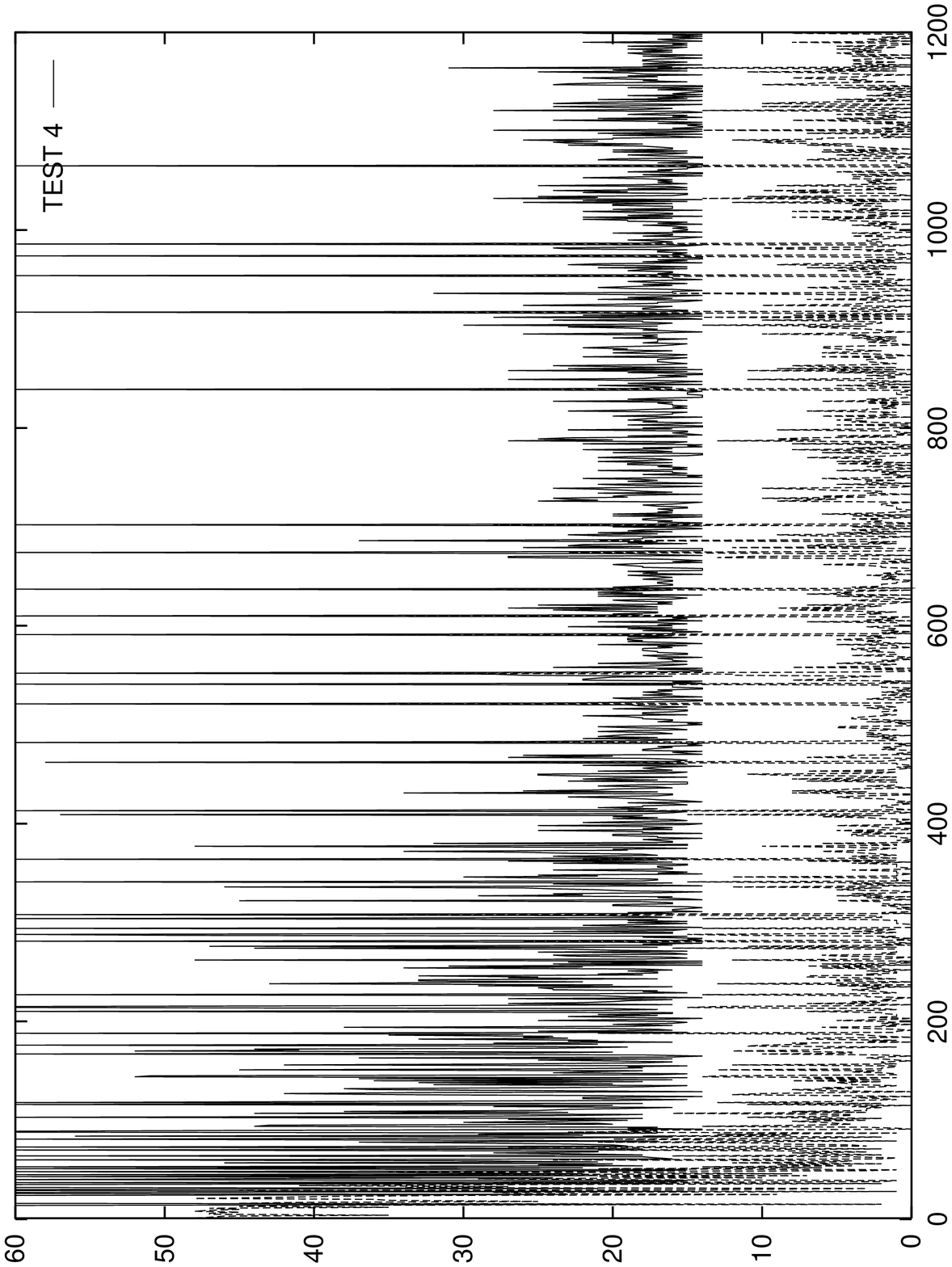}
\includegraphics{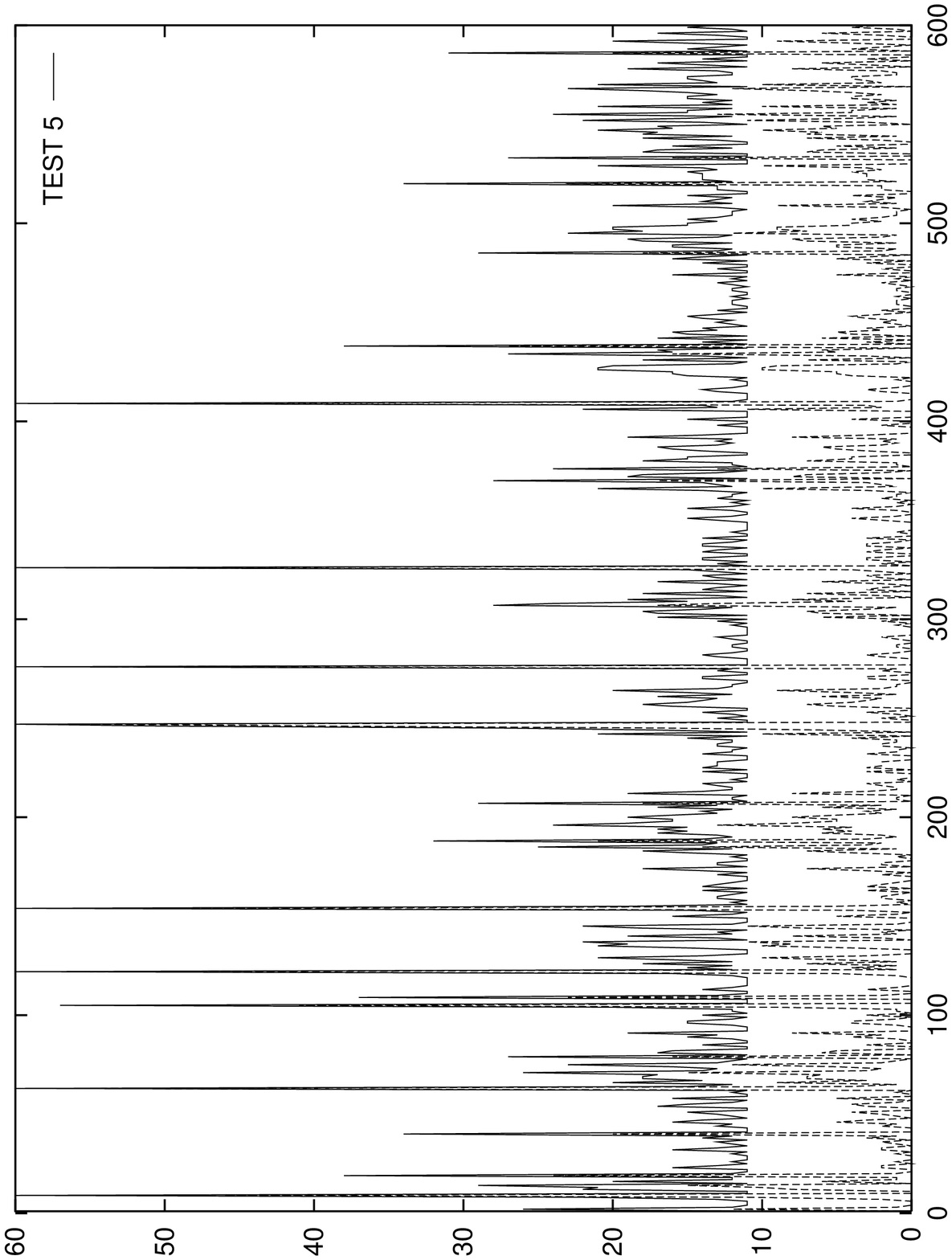}
\includegraphics{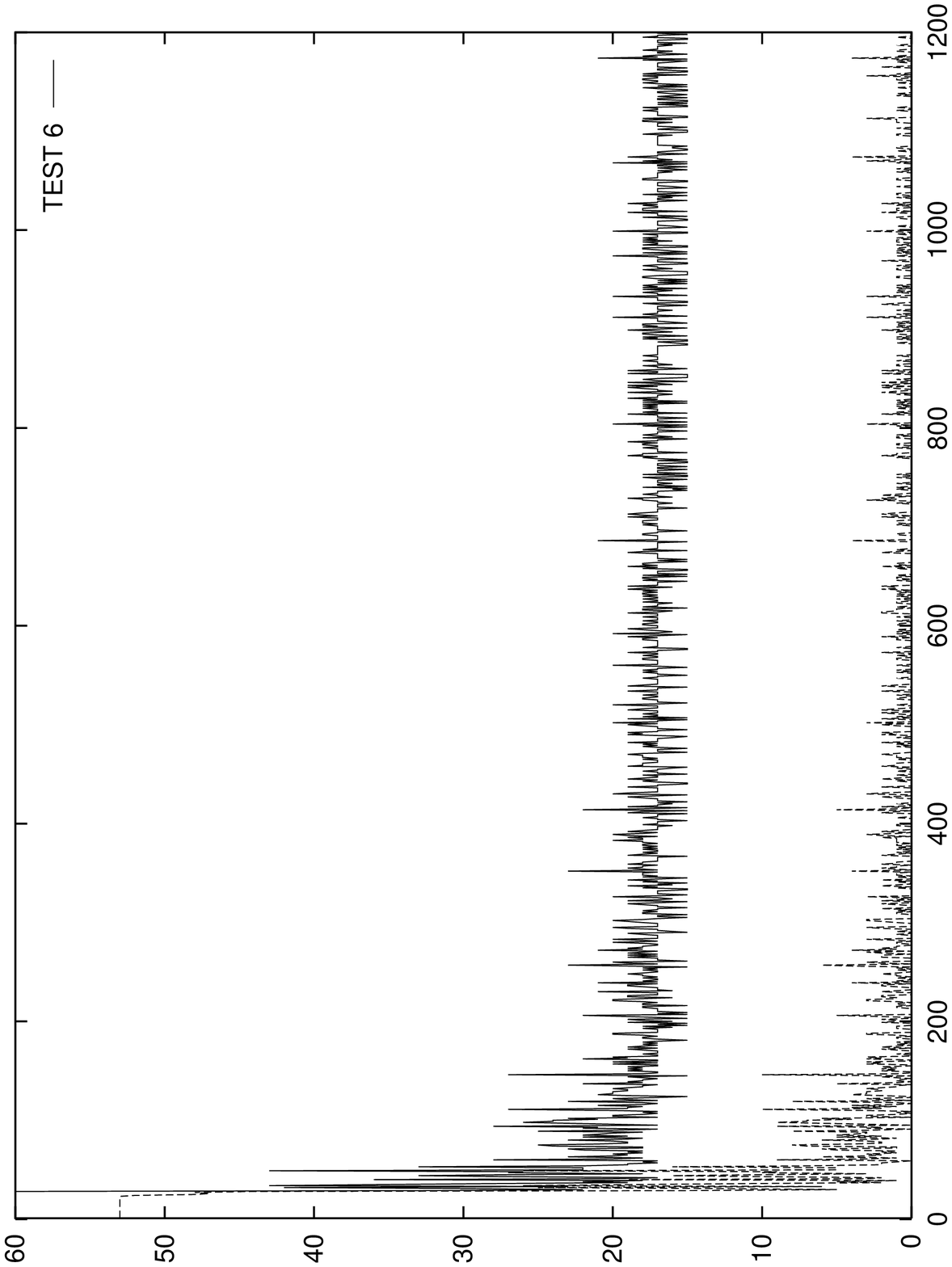}
\caption{{\it Performances for the 6 tests of Fig. \ref{f.arti1}. The full
line is the histogram of the total number of steps, including repeatedly 
running against obstacles (counted explicitly by the lower histogram -- 
dotted line). The parameters are not optimized}.
}
\label{f.arti2}
\end{figure}

\section{Conclusions}

The results described in section 2 indicate that the approach taken here
can be analized systematically and shows universal features. Notice that the 
parameter $a_1$ is of special significance: on the one hand it gives the 
interdependency between the particular steps during the learning period and on the other hand it turns out to be useful or even 
necessary for the convergence of the training. The ratio between the ``blind association" parameter $a_1$
and the ``unspecific
reinforcement" parameter $a_2$ does not appear arbitrary.

The illustration provided by the simulation of section 3 shows that the
elementary, primitive ``stochastic learning" considered here may cope also
with more ``differentiated" situations. This suggests that such mechanisms may well be of a rather fundamental nature. One can also ask about the
possibility of developing on this basis strategies for more evolved artificial
intelligence systems. 

\par\bigskip  

{\bf Acknowledgments:} The author thanks Reimer K\"uhn for agreeing to
present preliminary results from common work and for reading 
the manuscript. He is very indebted to 
Tai Tsun Wu for intensive and very helpful discussions.

\par\bigskip

\noindent {\bf References}\par\medskip

\no Barto, A.G., Sutton, R.S. and Anderson, Ch.W. (1983): 
``Neuronlike adaptive elements that can solve difficult learning 
control problems", IEEE Transactions on Systems, Man and Cybernetics, 
SMC-13, 834\par\medskip

\noindent
Hertz, J., Krogh, A. and Palmer, R.G. (1991) ``Introduction to the 
Theory of Neural Computation", Addison-Wesley, Reading (MA)
\par\medskip

\no Kaelbling, L.P., ed. (1996): Machine Learning. Special Issue on
Reinforcement Learning, Vol. 22, Nos. 1/2/3\par\medskip

\noindent K\"uhn, R. and Stamatescu, I.-O. (1998):``Statistical
Learning for Neural Networks", contribution to the workshop {\it
Fuzzy Logic, Fuzzy Control and Neural Networks}, ZiF, Bielefeld, 
April 7-11, 1997 and work in progress\par\medskip

\noindent Mlodinow, L. and
Stamatescu, I.-O. (1985): ``An evolutionary procedure for
machine learning",
{\it Int. Journal
of Computer and Inform. Sciences}, {\it 14}, 201.
\par\medskip

\no Sutton, R.S. (1988): ``Learning to predict by the method of temporal
differences", Machine Learning, 3, 9
\par\medskip

\noindent
Watkins, C.J.C.H. (1989): ``Learning from delayed rewards", Ph.D.Thesis

\end{document}